\documentclass[aps,prl,twocolumn,superscriptaddress,amsmath,amssymb,floatfix]{revtex4-2}
\usepackage{graphicx}
\usepackage{bm}
\usepackage{xcolor}

\usepackage[colorlinks=true,linkcolor=blue!60!black,citecolor=blue!60!black,urlcolor=blue!60!black]{hyperref}

\newcommand{\ii}{\mathrm{i}}
\newcommand{\Ecal}{\mathcal{E}}
\newcommand{\Acal}{\mathcal{A}}
\newcommand{\Lbar}{\bar{\mathcal L}}
\newcommand{\lmod}{\ell_{\mathrm{mod}}}
\newcommand{\leff}{\ell_{\mathrm{eff}}}
\newcommand{\ket}[1]{|#1\rangle}
\newcommand{\braket}[2]{\langle #1|#2\rangle}
\newcommand{\mel}[3]{\langle #1|#2|#3\rangle}

\newcommand{\slow}{\tilde{\nabla}}
\newcommand{\br}{\bm r}

\newcommand{\bk}{\bm k}
\newcommand{\bG}{\bm G}
\newcommand{\bR}{\bm R}
\newcommand{\bQ}{\bm Q}
\newcommand{\bQt}{\tilde{\bm Q}}
\newcommand{\bbeta}{\bm\beta}
\newcommand{\bbetat}{\tilde{\bm\beta}}
\newcommand{\balpha}{\bm\alpha}
\newcommand{\balphat}{\tilde{\bm\alpha}}
\newcommand{\bdelta}{\bm\delta}
\newcommand{\bdp}{\bm\delta'}
\newcommand{\ba}{\bm a}
\newcommand{\bv}{\bm v}
\newcommand{\bC}{\bm C}
\newcommand{\Lat}{\Lambda}
\newcommand{\Latt}{\tilde\Lambda}
\newcommand{\Mod}{\mathcal M}
\newcommand{\nuhat}{\hat\nu}
\newcommand{\vhat}{\hat{\bm v}}
\newcommand{\phat}{\hat{\bm p}}
\newcommand{\head}[1]{\textit{#1.---}}
\newif\ifshowptitles \showptitlesfalse      

\begin{document}

\title{Quasicrystals are Functionally Graded Materials}

\author{Michael J. Landry}
\email{mjlandry@mit.edu}
\affiliation{Quantum Measurement Group, MIT, Cambridge, MA 02139, USA}
\affiliation{Department of Nuclear Science and Engineering, MIT, Cambridge, MA 02139, USA}
\affiliation{Department of Physics, MIT, Cambridge, MA 02139, USA}

\author{Dominic Bollettieri}
\affiliation{Quantum Measurement Group, MIT, Cambridge, MA 02139, USA}
\affiliation{Department of Physics, MIT, Cambridge, MA 02139, USA}

\author{Akhil Sadam}
\affiliation{Center for Computational Science and Engineering, MIT, Cambridge, MA 02139, USA}

\author{Denisse Córdova Carrizales}
\affiliation{Quantum Measurement Group, MIT, Cambridge, MA 02139, USA}
\affiliation{Department of Nuclear Science and Engineering, MIT, Cambridge, MA 02139, USA}

\author{Mingda Li}
\email{mingda@mit.edu}
\affiliation{Quantum Measurement Group, MIT, Cambridge, MA 02139, USA}
\affiliation{Department of Nuclear Science and Engineering, MIT, Cambridge, MA 02139, USA}

\begin{abstract}
Quasicrystals exhibit long-range order without translational periodicity, posing challenges to compute their electronic properties. Here we show that a quasicrystal can be represented exactly as a periodic approximant with a spatially graded phason, establishing a connection to functionally graded materials. Building on this connection, we develop a controlled quantum theory of quasicrystals. We derive a state-dependent control parameter governed by the quantum metric of local Bloch states, determining when finite approximations are reliable. We further construct a Hamiltonian that correctly predicts Bragg gap positions, quasimomentum equivalence, and quasiperiodic eigenstates to all orders in phason strain within a finite Fourier window. We verify the theory numerically and propose a diffraction-based experimental test of the gap shift induced by phason strain. Our framework opens a route to predicting the electronic structure and functional properties of quasicrystals using methods developed for functionally graded materials.

\end{abstract}

\maketitle

Quasicrystals and functionally graded materials (FGMs) represent two distinct classes of spatially structured materials: Quasicrystals possess long-range order without translational periodicity~\cite{Shechtman1984,LevineSteinhardt1984}, with applications ranging from wear-resistant and low-friction coatings~\cite{Dubois2012} to prospective thermoelectric materials~\cite{Takagiwa2022} and engineered photonic structures~\cite{Vardeny2013}. FGMs, on the other hand, exhibit spatial variations in composition, microstructure, or orientation, enabling tailored properties for thermal-barrier coatings in aerospace and turbines~\cite{NiinoMaeda1990}, wear-resistant cutting tools~\cite{Cai2024}, and orthopedic implants~\cite{Sola2016}. Although traditionally treated as unrelated material families, both demonstrate how spatial organization can be exploited to control material properties.

Despite their technological importance, the spatial complexity also poses a fundamental challenge to theoretical predictions: the translational symmetry underlying conventional Bloch theory is absent in quasicrystals and generally broken by grading. For quasicrystals, a widely used approach replaces the aperiodic structure with a periodic approximant having a large unit cell, allowing conventional band-structure calculations~\cite{ElserHenley1985,GoldmanKelton1993,Fujiwara1989}. Alternative approaches include direct real-space calculations~\cite{VidarteLewenkopf2024}, higher-dimensional superspace formulations~\cite{DaviesThalhammer2025}, and quasiperiodic operator theory~\cite{JohnsonMoser1982,Bellissard1992,Jagannathan2021}. Recent developments have established spectral convergence bounds and certified calculations for infinite quasiperiodic systems~\cite{DaviesThalhammer2025,Chok2026}. However, a fundamental challenge remains in connecting the electronic states of practical periodic approximants to those of the underlying quasicrystal: how are their band energies, wavefunctions, and quantum numbers related, and when does a finite approximant faithfully describe them?

\begin{figure*}[t]
\includegraphics[width=\textwidth]{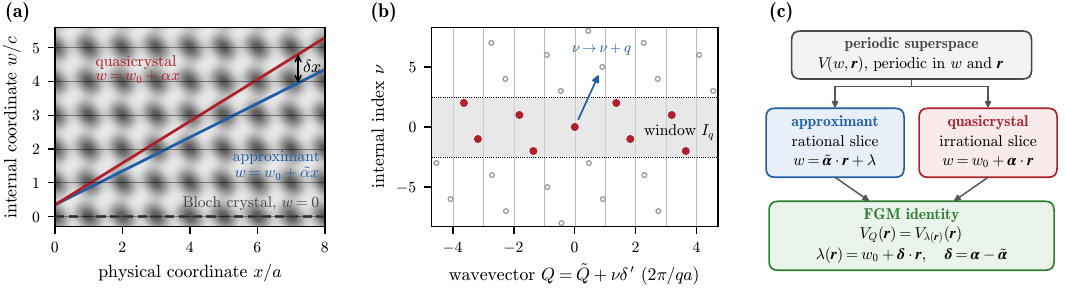}
\caption{A quasicrystal is a graded approximant. (a) A potential that is periodic in the physical coordinate $x$ and internal coordinate $w$ (gray scale). 
The Bloch crystal, quasicrystal, and approximant are the slices with slopes $w=0$, irrational $\alpha$, and nearby rational $\tilde\alpha=c/2a$, respectively, with $\xi=\alpha a/c$ the inverse golden mean.
The separation $\delta x$ between the last two, $\delta$ as in Eq.~\eqref{eq:FGM}, grows linearly in $x$; that is, the quasicrystal is the approximant with the drifting phason $w_0+\delta x$. (b) The Fourier module of the $3/5$ approximant in the plane of wavevector and internal index, with $\delta'$ exaggerated. Each supercell reciprocal vector $\tilde Q$ carries a column of quasicrystal wavevectors $\tilde Q+\nu\delta'$. The approximant keeps one point per column (filled), the one in the window $I_q$ of $q$ consecutive internal indices, $-q/2<\nu\le q/2$ (gray band), and identifies $\nu$ with $\nu+q$. Our theory moves each filled point to its true position, but it cannot reach the open ones. (c) The theory of functionally graded materials can be applied to the phasons of the quasicrystal and its approximant. The FGM parameter $\lambda$ becomes defined in terms of $\delta x$, which varies slowly on the scale of the lattice constant, though not on the scale of the supercell (see \emph{The small parameter}).}
\label{fig:one}
\end{figure*}

In this Letter, we establish a connection between quasicrystals and FGMs, enabling a controlled quantum theory to study quasicrystal electronic structure. We show that a quasicrystal can be represented exactly as a periodic approximant with a spatially graded phason, arising from the difference between irrational and rational cuts of a higher-dimensional crystal~\cite{deWolff1974,JannerJanssen1977,KaluginKitaevLevitov1985,DuneauKatz1985} (Fig.~\ref{fig:one}). Building on the recent quantum theory of FGMs~\cite{Landry2026}, we derive a state-dependent control parameter governed by the quantum metric of local Bloch states, establishing when finite approximations are reliable and when they fail. We further construct a Hamiltonian that correctly predicts Bragg gap positions, quasimomentum equivalence, and quasiperiodic eigenstates to all orders in phason strain within a finite Fourier window. Our framework thus transforms periodic approximants into a controlled computational approach to quasicrystals, opening a route to predicting their electronic structure and functional properties beyond conventional band theory.

\head{The quasicrystal as a graded approximant}
We consider a minimal representative quasicrystal with one phason, and show it can be represented exactly as a spatially varying phason modulation of a periodic approximant~\footnote{We consider the single phason model for simplicity, given by a 4D embedding; generically there may be up to 3 phasons in a 6D embedding. Generalizing the results of this Letter is straightforward.}.
Denote the quasicrystal potential by $V_Q(\br\in\mathbb R^{3})$, defined as the restriction of a higher-dimensional periodic potential $V(w,\br)$ over $\mathbb R^{4}$ to a 3D hyperplane, with discrete symmetries $w\to w+c$ and $\br \to \br+\bR_i$ for basis $\bR_i$ of lattice $\Lat$ with reciprocal lattice $\Lat^{*}$.  Defining $\bG$ as the wavevector and $n$ as the superspace harmonic, we have $V=\sum_{n\bG}V_{n\bG}e^{2\pi\ii nw/c}e^{\ii\bG\cdot\br}$. 
The quasicrystal potential is then
\begin{equation}
V_Q(\br)=V(w_0+\balpha\!\cdot\!\br,\br)=\sum_{n\bG}V_{n\bG}\,e^{2\pi\ii nw_0/c}\,e^{\ii(n\bbeta+\bG)\cdot\br},
\label{eq:VQ}
\end{equation}
with $\balpha$ and $w_0$ the hyperplane slope and intercept (the \emph{phason}) respectively, and wavevector $\bQ=n\bbeta+\bG$ with $\bbeta=2\pi\balpha/c$. 
These wavevectors form a Fourier module $\Mod=\Lat^{*}+\mathbb Z\bbeta$, which are dense in $\mathbb R^{3}$ whenever $1$ and the hyperplane slopes $\xi_i=\balpha\!\cdot\!\bR_i/c$ are rationally independent~\cite{Lifshitz2003}; in one dimension, $\xi=\alpha a/c$ reduces to a period ratio. 
The decomposition $\bQ=n\bbeta+\bG$ is then unique in $n$, so $n$ is a function $\nu(\bQ)$ on the module, which we define as the \emph{internal harmonic index}. The corresponding operator $\nuhat$ scales plane waves of form $e^{\ii(\bk+\bQ)\cdot\br}$ with Bloch wavevector $\bk$ by $\nu(\bQ)$,
and is a generator of phason translations via the unitary shift operator $e^{2\pi\ii\nuhat w_0/c}$, which preserves the energy spectrum~\cite{Bak1985PRL,Baggioli:2020haa}. 

Supercells approximate $\balpha$ by a rational slope $\balphat$, where $\balphat\!\cdot\!\bR_i/c=p_i/q$. The supercell then holds $q$ cells. The supercell's lattice is $\Latt=\{\bR\in\Lat:\balphat\!\cdot\!\bR\in c\mathbb Z\}$ with reciprocal lattice $\Latt^{*}=\Lat^{*}+\mathbb Z\bbetat$ (vectors $\bQt$), $\bbetat=2\pi\balphat/c$, and its Fourier module is the same but $\nu$ has been reduced modulo $q$.

%
We now cast the quasicrystal potential into an approximant. Let $V_\lambda(\br)\equiv V(\balphat\!\cdot\!\br+\lambda,\br)$ for some $\lambda$. Then
 \begin{equation}\begin{gathered}
 V_Q(\br)= V(w_0+\bdelta\!\cdot\!\br+\balphat\!\cdot\!\br,\br) = V_{\lambda(\br)}(\br),\\
\lambda(\br)\equiv w_0+\bdelta\!\cdot\!\br,\quad\bdelta\equiv\balpha-\balphat ,
\label{eq:FGM}
\end{gathered}\end{equation}
where we have parameterized the set of approximants by a spatially varying phason, exactly reproducing an FGM potential $V_{\lambda(\br)}(\br)$~\cite{Landry2026}.
Previous diffraction experiments have measured $\bdelta$, the phason strain, which displaces every diffraction peak of internal index $\nu$ by $\nu\bdp$, with $\bdp\equiv2\pi\bdelta/c$ its wavevector~\cite{Lubensky1986,Ishii1992}. 

%

\head{Two key observations}
(1) Translating $V_\lambda$ by a lattice vector in $\Lambda$  yields $V_{\lambda+\balphat\cdot\bR}$, where $\balphat\cdot\bR$ is a multiple of $c/q$. Let $\Ecal_n(\bk;\lambda)$ be the bands of $V_\lambda$ and $u_{n\bk}(\br;\lambda)$ the periodic parts of its Bloch functions, normalized over the supercell. 
As band structure is translation-invariant,
\begin{equation}
\Ecal_n(\bk;\lambda+c/q)=\Ecal_n(\bk;\lambda).
\label{eq:period}
\end{equation}
%
Thus the local band structure must repeat $q$ times per phason period, with an effective modulation length $\leff=\lmod/q$ for a phason modulation length $\lmod=c/|\bdelta|$. 

(2) What does shifting the phason $\lambda$ do to the approximant potential? To answer this question, notice that for the superlattice, $\nu(\bQt)$ is not unique but is defined only modulo $q$; we therefore 
restrict $\nu(\bQt)\in I_q$, a window of $q$ consecutive integers, e.g., $-q/2<\nu\le q/2$. Notice that the harmonics of $V$ connecting $\bQt$ to $\bQt'$ have internal index $\nu(\bQt)-\nu(\bQt')+\ell q$, $\ell\in\mathbb Z$. 
For the $\ell=0$ terms, potentials of different $\lambda$ are connected by a unitary transformation $U_\lambda\equiv e^{2\pi\ii\nuhat\lambda/c}$. We account for the $\ell\neq 0$ terms in a wrap-around potential, denoted $\tilde V_\lambda$, which contains high internal harmonics of $V$ and couplings across the two edges of the window, resulting in
\begin{equation}
V_\lambda=U_\lambda V_0U_\lambda^{\dagger}+\tilde V_\lambda .
\label{eq:wrap}
\end{equation}

$U_\lambda$ is unitary, so all phason-modulated band structure effects originate from $\tilde V_\lambda$.
As $\bQ=\bQt+\nu\bdp$ [Fig.~\ref{fig:one}(b)], the approximant is the quasicrystal with two operations applied: $\nu\sim\nu+q$, and $(\bk+\bQt+\nu\bdp)^{2}\to(\bk+\bQt)^{2}$ in the kinetic energy.
The gradient expansion undoes the latter. The former, unfortunately, cannot be undone; the approximant is only accurate up to the supercell window. We will return to this point when we define a window Hamiltonian.

\head{The small parameter}
For a supercell of size $qa$, there are two naive candidate parameters.
We might expect the small parameter to be either $\kappa_0=qa/\lmod=q|\xi-p/q|$ or $\kappa=qa/\leff=q^{2}|\xi-p/q|$, based on the irrational hyperplane slope.
We can bound the lowest possible error of such an approximant, taking $\xi$ to be the golden mean for example. Using a continued fraction expansion with convergents $p_n/q_n$ for a supercell of $q_n$ cells, 
\cite{HardyWright2008,Khinchin1997} show that $\kappa_n\equiv q_n^{2}|\xi-p_n/q_n|$ is bounded below by $1/(a_{n+1}+2)$ for $a_{n+1}$ the next partial quotient. By Hurwitz's theorem~\cite{Hurwitz1891} we find  $\kappa_n$ converges to $1/\sqrt5$.
There is therefore a trade-off between a small gradient and a small supercell: the band structure of a golden-mean approximant repeats every $1/\kappa_n\approx2.24$ supercells, irrespective of the supercell size.
A similar argument holds for any quadratic irrational, and observed quasicrystals likewise have $\xi$ quadratic irrational~\cite{Levitov1988}.

It turns out 
that neither $\kappa_0$ nor $\kappa$ is the correct small parameter for all states. When the wrap-around potential is negligible, Eq.~\eqref{eq:wrap}
reduces to $u(\lambda)=U_\lambda u(0)$ for the periodic part of the Bloch function, with slow derivative
\begin{equation}
\slow u\equiv\bdelta\,\partial_\lambda u =\ii\,\bdp\,\nuhat\,u .
\label{eq:slowu}
\end{equation}
Each plane-wave component accumulates phase (winds) proportional to the internal index, via $e^{\ii\nu\bdp\cdot\br}$: components with $\nu=\pm1,\pm q/2$ wind once over $\lmod,\leff$ respectively. 
As a representative example, we let $\bdp \propto \hat{\bm e}_z$, for which the Hilbert-space distance between the Bloch states of neighboring points, $1-|\braket{u(z)}{u(z{+}dz)}|^{2}=g_{zz}dz^{2}$, with $g_{zz}$ the quantum metric, is therefore
\begin{equation}
g_{zz}=|\bdp|^{2}\Delta\nu^{2},\qquad\Delta\nu^{2}\equiv\langle\nuhat^{2}\rangle-\langle\nuhat\rangle^{2},
\label{eq:metric}
\end{equation}
with $\langle\cdot\rangle$ the expectation value in $u$.
%
A natural small parameter can then be defined in terms of this metric: $\kappa_{\mathrm{eff}}=qa\sqrt{g_{zz}}/2\pi=\kappa_0\Delta\nu$.
As shown in Ref.~\cite{Landry2026}, a slow theory can be obtained by averaging the action over a supercell, and such an average discards supercell harmonics that have slow phases $e^{\ii(\nu-\nu')\bdp\cdot\br}$. Each such term can be integrated by parts, each time gaining a factor of $\kappa_0|\nu-\nu'|$. 
The average is therefore exact up to terms that wind by $\kappa_0\Delta\nu$ per supercell. In particular, then, an approximant is a good approximation to the quasicrystal depending on both the size of the supercell and on the state of interest. For example, a state comprising a few plane waves has $\Delta\nu$ of order one and $\kappa_{\mathrm{eff}}\to0$ with $q$; a state spread over the supercell window has $\Delta\nu\sim q$ and $\kappa_{\mathrm{eff}}\simeq\kappa$. 
%

\begin{figure*}[t]
\includegraphics[width=\textwidth]{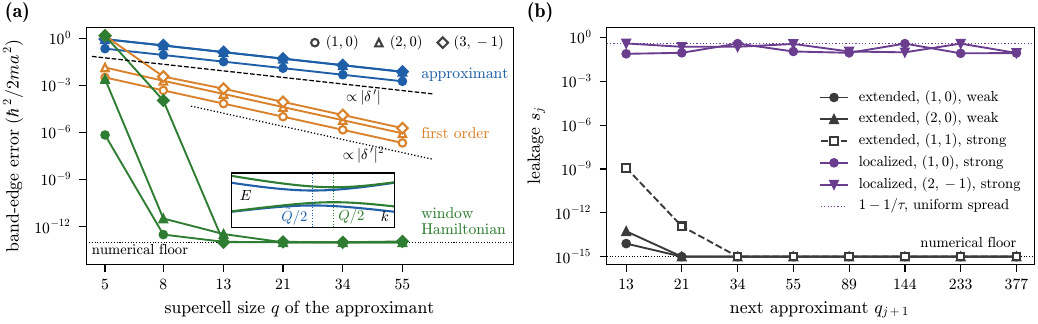}
\caption{Numerical test on the one-dimensional potential $2W_0\cos(2\pi x/a)+2W_1\cos(2\pi\xi x/a+\varphi)$, with $\xi$ the inverse golden mean and $p/q$ its Fibonacci approximants. (a) Weak coupling, $W_0=0.3$ and $W_1=0.5$ in $\hbar^{2}/2ma^{2}$ units: lower edge error of three gaps, relative to the $233/377$ window Hamiltonian, as computed from the approximant (filled blue), the first-order correction of Eq.~\eqref{eq:E2} (open orange), and the window Hamiltonian of Eq.~\eqref{eq:window} (green). The dashed and dotted lines are proportional to $|\delta'|\propto1/q^{2}$ and $|\delta'|^{2}$. The window Hamiltonian is exact
except where the state has nonnegligible weight at the window edge, as seen at $q=5, 8$.
Inset: predicted gap $(1,0)$ positions by the approximant and window Hamiltonian at $\tilde Q/2$ and $Q/2$ respectively. (b) Leakage $s_j$, the weight of a state in the shell $q_j/2<|\nu|\le q_{j+1}/2$ that the approximant $q_j$ omits and the next convergent retains. Extended states (gray) reach the numerical floor within two convergents. The two low-lying states at strong coupling, $W_0=3$ and $W_1=5$ (purple), are localized in a few wells of the potential, and their leakage never decreases; the purple dotted line, $1-1/\tau$ with $\tau$ the golden ratio, is the leakage of a state spread uniformly over the window.}
\label{fig:two}
\end{figure*}

\head{The complete slow theory}
We start from the modulated Bloch state of Ref.~\cite{Landry2026}, $\psi=u_{n\bk(\br)}(\br;\lambda(\br))\,\Acal e^{\ii S}$, with $\Acal$ a slowly varying amplitude and $\bk=\nabla S$ the slow lattice momentum. Inserting $\psi$ into the Schr\"odinger action and averaging over a supercell gives a Lagrangian for the slow fields with at most two slow derivatives~\cite{Landry2026},
\begin{equation}
\Lbar=-\Acal^{2}\big[\Ecal_n(\bk{+}\ba)+\delta\Ecal_n+U_g-E\big]-\tfrac12\nabla\Acal\!\cdot\!\mathbb M^{-1}\!\cdot\!\nabla\Acal ,
\label{eq:action}
\end{equation}
%
exact to all orders in $\kappa_{\mathrm{eff}}$. Here $\bk+\ba$ is the gauge-invariant momentum, with $\ba=\mel{u}{\tfrac1\ii\slow}{u}$ the Berry connection of the local Bloch states~\cite{Landry2026}, a pseudo vector potential; $\delta\Ecal_n=\mathrm{Im}\mel{u}{\vhat_i(1-P_n)}{\slow_iu}$ is the gradient energy of Sundaram and Niu~\cite{SundaramNiu1999}, with $\vhat=\phat/m$ the velocity, $\phat\ket{\bk+\bQt}=(\bk+\bQt)\ket{\bk+\bQt}$, and $P_n$ the band projector; $U_g=\mathrm{tr}\,(g/2m)$ is the quantum-metric potential; $\mathbb M^{-1}=\partial^{2}\Ecal_n/\partial\bk\partial\bk$ is the inverse band mass. For a generic graded crystal these are unwieldy sums over bands. For a quasicrystal, though, we can use Eq.~\eqref{eq:slowu} 
as $\vhat$ and $\nuhat$ are both diagonal in the plane-wave basis:
\begin{equation}
\ba=\bdp\langle\nuhat\rangle,\quad\delta\Ecal_n=\bdp\!\cdot\!\mathrm{Cov}(\vhat,\nuhat),\quad U_g=\frac{\bdp^{2}\Delta\nu^{2}}{2m},
\label{eq:three}
\end{equation}
with $\mathrm{Cov}(\vhat,\nuhat)=\sum_{\bQt}(\nu-\langle\nuhat\rangle)(\bk+\bQt)|u_{\bQt}|^{2}/m$ the correlation between the velocity of a plane-wave component and its internal index and $u_{\bQt}$ the plane-wave coefficients of $u$. Expanding Eq.~\eqref{eq:action} about 
$\bk$ (equivalently, applying second-order perturbation theory to the window Hamiltonian below), we find that the energy of the state $(n,\bk)$ through second order is
\begin{equation}
E_{n\bk}=\Ecal_n+\bdp\!\cdot\!\bC_n+\frac{\bdp^{2}\langle\nuhat^{2}\rangle}{2m}-\sum_{n'\neq n}\frac{|\bdp\!\cdot\!\bC_{n'n}|^{2}}{\Ecal_{n'}-\Ecal_n},
\label{eq:E2}
\end{equation}
with $\bC_{n'n}\equiv\mel{u_{n'}}{\tfrac12\{\vhat,\nuhat\}}{u_n}$ and $\bC_n=\bC_{nn}=\langle\nuhat\rangle\bv_n+\mathrm{Cov}(\vhat,\nuhat)$, $\bv_n=\partial_{\bk}\Ecal_n$ the band velocity. Equation~\eqref{eq:E2} holds if $|\bdp\!\cdot\!\bC_{n'n}|\ll|\Ecal_{n'}-\Ecal_n|$; otherwise the bands must be diagonalized, which we consider in Eq.~\eqref{eq:window}. The first-order term is the Hellmann-Feynman shift produced by moving every plane-wave component of the approximant's Bloch state from $\bk+\bQt$ to its true position $\bk+\bQt+\nu\bdp$. 
At the band edge, the state is an equal superposition of two plane waves whose internal indices differ by $n_\perp$, for a module wavevector $\bQ=n_\perp\bbeta+\bG$. Thus the band velocity, and in turn $\ba\!\cdot\!\bv_n$ are both 0. The covariance term $\bdp\!\cdot\!\mathrm{Cov}(\vhat,\nuhat)$ is $n_\perp\bdp\!\cdot\!\bQt/4m$ with $\bQt=\bQ-n_\perp\bdp$. For a simple (ungraded) approximant, the wavevector is $\bQt$, so the Bragg crossing is predicted to be $\bQt/2=\bQ/2-n_\perp\bdp/2$ instead of $\bQ/2$. As $\ba\!\cdot\!\bv_n = 0$, retaining the Berry connection does not improve the Bragg crossing.

\head{Eigenstates, quantum numbers, band structure}
When the wrap-around potential is negligible, the local band structure does not depend on the phason, so the equations of motion of Eq.~\eqref{eq:action}, $\Ecal_n(\bk{+}\ba)+\delta\Ecal_n+U_g=E$ and $\nabla\!\cdot\!(\Acal^{2}\bv_n)=0$, have constant solutions. For a volume $\mathcal V$, the eigenstate is then
\begin{equation}
\psi_{n\bk}=\frac{1}{\sqrt{\mathcal V}}\sum_{\bQt\in\Latt^{*}}u_{n\bk,\bQt}\,e^{2\pi\ii\nu(\bQt)w_0/c}\,e^{\ii(\bk+\bQt+\nu(\bQt)\bdp)\cdot\br}.
\label{eq:state}
\end{equation}
Its wavevectors $\bk+\bQt+\nu\bdp=\bk+n\bbeta+\bG$ are exactly the module points $\bk+\Mod$. Equation~\eqref{eq:state} is therefore simply the plane-wave superposition solution from KAM theory for the extended states of a weak quasiperiodic potential~\cite{DinaburgSinai1975,Eliasson1992}, with the approximant supplying coefficients. As the approximant's state is periodic, only the phases $e^{\ii\nu\bdp\cdot\br}$ make Eq.~\eqref{eq:state} quasiperiodic, so the corresponding momentum observable by photoemission would be at $\hbar(\bk+\bQ)$ rather than $\hbar(\bk+\bQt)$. 
The quantum numbers are the supercell band index $n$ and Bloch wavevector $\bk$. We quickly show the quasicrystal band structure is correctly periodic under module $\Mod$.
The approximant identifies $(n,\bk) \sim (n,\bk+\bQt)$ as the same state, as $u_{n,\bk+\bQt}=e^{-\ii\bQt\cdot\br}u_{n\bk}$, but $e^{-\ii\bQt\cdot\br}$ also
shifts the internal indices by $\nu(\bQt)$, and by Eq.~\eqref{eq:slowu} induces a slow phase on $u$, so $\bk$ shifts by $\nu(\bQt)\bdp$. Thus, $(n,\bk) \sim( n,\bk+\bQt; \nu \rightarrow \nu + \nu(\bQt)) \sim( n,\bk+\bQ)$ with $\bQ\in\Mod$.
We now consider band gaps from a k-space viewpoint. 
In one dimension, a gap corresponding to a module wavevector $n_\perp\beta+2\pi m/a$ can be identified by the internal and lattice harmonics, $(n_\perp,m)$, and by the gap-labeling theorem~\cite{JohnsonMoser1982,AvronSimon1983,Bellissard1992}, the integrated density of states converges correctly regardless of the choice of approximant $|n_\perp p+mq|/qa\to|n_\perp\xi+m|/a$. 
Equation~\eqref{eq:E2} moves only the positions of the gaps, not their labels. Finally, we have assumed each gap is resolved by a single energy band. This condition requires the gap to be wider than the maximum possible phason-induced displacement, $\Delta_{\mathrm{gap}}\gg|n_\perp||\bdp||\bv|$, with $\bv=\bQt/2m$ the velocity at the Bragg plane. As $\Delta_{\mathrm{gap}}$ shrinks geometrically with $|n_\perp|$, any finite size approximant  can resolve at most finitely many gaps, and will necessarily miss the Cantor-set-like fine structure. 

\head{All orders}
Thus far, we have treated Eq.~\eqref{eq:E2} as a perturbative expansion. We now show it is an expansion of the supercell window Hamiltonian.
Consider the subspace $\mathcal S_{\bk}$ spanned by the window module wavevectors $\bk+\bQt+\nu(\bQt)\bdp$. The interior potential consists of the $\ell=0$ terms of Eq.~\eqref{eq:wrap}, so
\begin{equation}
H_Q\big|_{\mathcal S_{\bk}}=U_{w_0}\tilde H_{\bk}U_{w_0}^{\dagger},\quad
\tilde H_{\bk}=H_0(\bk)+\frac{\bdp\!\cdot\!\{\phat,\nuhat\}+\bdp^{2}\nuhat^{2}}{2m},
\label{eq:window}
\end{equation}
\emph{exactly}, with $H_Q=-\nabla^{2}/2m+V_Q$ the true Hamiltonian and $H_0(\bk)$ the approximant Hamiltonian at $\lambda=0$. The FGM action is the perturbation theory of this window Hamiltonian $H_Q\big|_{\mathcal S_{\bk}}$,
with $\tilde H_{\bk}$ quadratic in the strain.
But $\tilde H_{\bk}$ is just the approximant's plane-wave Hamiltonian with each kinetic energy moved to $(\bk+\bQt+\nu\bdp)^{2}/2m$,
as the wrap-around potential does not act within $\mathcal S_{\bk}$.
To solve for the true eigenstates, we instead diagonalize $\tilde H_{\bk}$ and thus $H_Q\big|_{\mathcal S_{\bk}}$ to all orders of $\bdp$, with the only remaining error due to the window $\mathcal S_{\bk}$.
Diagonalizing $\tilde H_{\bk}$ is no more difficult than diagonalizing the approximant's Hamiltonian, and allows computation of eigenstates by inserting the eigenvectors into Eq.~\eqref{eq:state}. The spectrum is also correctly phason-invariant, as $w_0$ enters only through the unitary $U_{w_0}$.
%
Now consider the error due to $\mathcal S_{\bk}$. Assuming an extended state, such that the coefficients decay with the internal index on a scale $\zeta$, the error is of order $e^{-q/2\zeta}$. Physically, the local band structure repeats after a phason advance of $c/q$, Eq.~\eqref{eq:period}. The local Bloch state therefore has degenerate copies of $u$ with internal indices shifted by $jq$, carrying slow momentum $jq\bdp$, by Eq.~\eqref{eq:slowu}. These copies place the plane waves at $\bQt+(\nu+jq)\bdp$, the open points of Fig.~\ref{fig:one}(b), and since $jq|\bdp|\leff=2\pi j$ they never separate from the state itself.
The wrap-around potential is simply the coupling to these degenerate copies of $u$: its $\ell$th term carries phase $e^{2\pi\ii\ell q\lambda/c}$, which at $\lambda(\br)$ becomes the slow phase $e^{\ii\ell q\bdp\cdot\br}$, scattering the slow momentum by $\ell q\bdp$. 
For a window Hamiltonian $(2J{+}1)q$ wide, we can couple copies with $|j|\le J$. 
Therefore enlarging the supercell window is equivalent to coupling degenerate local Bloch states. We now describe ways to compute extended, localized, and critical states. Critical states are neither extended nor localized, decaying as a power law rather than exponentially. For an extended state, a small finite supercell suffices, with a weight on the $j$th Bloch state proportional to $e^{-(2|j|-1)q/2\zeta}$. For a localized state, coefficients are spread uniformly over the internal index, so no finite window captures it [Fig.~\ref{fig:two}(b)], though in real space a supercell that includes the environment is sufficient. For a critical state, we require sufficient resolution in both $\nu$ and $\br$, so those states are not directly computable. A possible indirect approach would treat critical states as fixed points of the process of enlarging the supercell window; we return to this point in the discussion.

\head{Numerical tests}
We test the theory in one dimension under a bichromatic potential (Figure~\ref{fig:two}), which is an example continuum model of cold atoms in two incommensurate optical lattices~\cite{Roati2008,Modugno2009,Luschen2018}. At weak coupling, the approximant's band-edge error is $n_\perp\delta'\tilde Q/4m$, decreasing as $|\delta'|\propto1/q^{2}$; after first-order correction [Eq.~\eqref{eq:E2}] the residual is $\propto|\delta'|^{2}$. The Berry connection component $\ba\!\cdot\!\bv$ does not affect the edges (to $10^{-6}$); and finally the window Hamiltonian, at its own extremum over $k$, is accurate to better than $10^{-12}$ for all three gaps from $q=13$ on. Here, we use the window Hamiltonian of the $233/377$ approximant as the reference solution, whose only error is its own leakage.
One might worry that we are merely comparing the window Hamiltonian with itself. 
To rule this out, we also computed the same band edges by diagonalizing $H_Q$ directly on the module points $k+2\pi(m+n\xi)/a$ with $|n|\le40$ and kinetic energy below $400\,\hbar^{2}/2ma^{2}$, about $530$ plane waves, with no approximant and no folding; the two references agree to $2\times10^{-13}$, the numerical floor of Fig.~\ref{fig:two}(a).
Checking Eq.~\eqref{eq:period}, the approximant's energy spectra at phason phase $\varphi=2\pi w_0/c$ and $\varphi+2\pi/q$ agree to $10^{-13}$, with sub-period variation decaying as $e^{-1.8q}$. At strong coupling, the same potential can host both extended and localized states: high-level states have $\Delta\nu\simeq0.5$ and behave as if at weak coupling, while low-lying states are localized in a few wells of the potential, with $\Delta\nu\simeq0.29q\simeq q/\sqrt{12}$, and are not captured due to state leakage [Fig.~\ref{fig:two}(b)]. Critical states occur in this potential only at a mobility edge, which would require careful tuning~\cite{AubryAndre1980,BiddleDasSarma2010}, so we do not consider them.

\head{Discussion}
We have shown that a quasicrystal can be represented exactly as a phason-graded periodic approximant [Eq.~\eqref{eq:FGM}], establishing a connection to functionally graded materials. This viewpoint provides a controlled framework for calculating quasicrystal electronic structure, including band energies, quantum numbers, and quasiperiodic eigenstates, while identifying which spectral features can be reliably resolved by finite approximations.

In particular, our framework yields the correct wavevector equivalence $(n,\bk)\sim(n,\bk+\bQ)$, reconstructs quasiperiodic eigenstates [Eq.~\eqref{eq:state}], and corrects Bragg crossing wavevectors by $n_\perp\bdp/2$, predicting photoemission momenta at $\hbar(\bk+\bQ)$ rather than $\hbar(\bk+\bQt)$. The resulting window Hamiltonian gives band energies to all orders in phason strain within a finite Fourier window, while preserving spectral invariance under phason translations. Furthermore, the quantum-metric control parameter establishes when a finite approximation is reliable and identifies unresolved fine spectral structures with $\Delta_{\mathrm{gap}}\not\gg|n_\perp||\bdp||\bv|$. These advances provide a foundation for predicting not only the electronic structure of quasicrystals, but also their spectroscopic, optical, and transport properties through extensions of existing response theories.

Our framework also suggests new opportunities for understanding and engineering quasiperiodic matter. In particular, the explicit connection between phason modulation and electronic structure offers a route to exploring how spatial variations in quasiperiodic order influence electronic properties, potentially enabling the design of functional quasicrystalline structures. For localized and critical states, where the control parameter becomes large, further developments are required. We conjecture that critical states may be understood as fixed points of a renormalization procedure involving successively larger Fourier windows. Another important direction is extending the theory to metallic quasicrystals, where discontinuous atomic rearrangements under phason variation challenge the present smooth-potential assumption.

Finally, we propose a direct experimental test connecting phason strain to electronic structure. The phason strain can be measured through the displacement $\nu\bdp$ of diffraction peaks, while our theory predicts a corresponding Bragg-crossing energy shift of $n_\perp\bdp\!\cdot\!\bv/2$, with $\bv=\bQt/2m$, relative to the periodic approximant. This relation could be tested by combining diffraction with electronic spectroscopy in quasicrystals~\cite{Edagawa2004}, or through analogous measurements in detuned bichromatic optical lattices and moir\'e quasicrystals~\cite{Ahn2018,Yao2018,Uri2023}. More broadly, our work extends the computational framework of functionally graded materials to quasiperiodic matter, opening possibilities for predictive modeling and ultimately the design of quasicrystals with tailored quantum properties.

\begin{acknowledgments}
The authors acknowledge support from the National Science Foundation (NSF) Convergence Accelerator Award No.~2345084 and the
U.S. Department of Energy (DOE), Basic Energy Sciences (BES), Award No. DE-SC0020148. D.C.C acknowledges support from the Jack Kent Cooke Foundation and The Paul and Daisy Soros Fellowships for New Americans. M.~L. acknowledges support from R.~Wachnik.
\end{acknowledgments}

\bibliographystyle{apsrev4-2}
\bibliography{references}

\end{document}